\documentclass[prb,twocolumn,aps,superscriptaddress]{revtex4-2}

\usepackage{graphicx}
\usepackage{dcolumn}
\usepackage{bm}
\usepackage{float} 
\usepackage{comment}
\usepackage{amssymb}

\usepackage{amsmath}
\usepackage{physics}
\usepackage{sublabel}
\usepackage[utf8]{inputenc}
\usepackage{hyperref}
\usepackage[usenames, dvipsnames]{color}
\usepackage[english]{babel}
\usepackage[T1]{fontenc}
\usepackage{bbm}

\usepackage{float}
\hypersetup{colorlinks = true, linkcolor=blue, citecolor=blue, urlcolor=blue}

\begin{document}

 \author{Siddhant Midha}
 \affiliation{Department of Electrical Engineering, Indian Institute of Technology Bombay, Powai, Mumbai-400076, India
}

\author{Koustav Jana}
 \altaffiliation[Current Address: ]{Department of Applied Physics, Stanford University, Stanford, CA 94305, USA.}
 \affiliation{Department of Electrical Engineering, Indian Institute of Technology Bombay, Powai, Mumbai-400076, India
}
\author{Bhaskaran Muralidharan}%
\email{bm@ee.iitb.ac.in}
\affiliation{Department of Electrical Engineering, Indian Institute of Technology Bombay, Powai, Mumbai-400076, India
}%

\title{Are symmetry protected topological phases immune to dephasing?}

\begin{abstract}
Harnessing topological phases with their dissipationless edge-channels coupled with the effective engineering of quantum phase transitions is a spinal aspect of topological electronics. The accompanying symmetry protection leads to different kinds of topological edge-channels which include, for instance, the quantum spin Hall phase, and the spin quantum anomalous Hall phase. To model realistic devices, it is important to ratify the robustness of the dissipationless edge-channels, which should typically exhibit a perfect quantum of conductance, against various disorder and dephasing. This work is hence devoted to a computational exploration of topological robustness against various forms of dephasing. For this, we employ phenomenological dephasing models under the Keldysh non-equilibrium Green's function formalism using a model topological device setup on a 2D-Xene platform. Concurrently, we also explicitly add disorder via impurity potentials in the channel and averaging over hundreds of configurations. To describe the extent of robustness, we quantify the decay of the conductance quantum with increasing disorder under different conditions. 
Our analysis shows that these topological phases are robust to experimentally relevant regimes of momentum dephasing and random disorder potentials. We note that Rashba mixing worsens the performance of the QSH phase and point out a mechanism for the same.  Further, we observe that the quantum spin Hall phase break downs due to spin dephasing, but the spin quantum anomalous Hall phase remains robust. The spin quantum anomalous Hall phase shows stark robustness under all the dephasing regimes, and shows promise for realistic device structures for topological electronics applications.
\end{abstract}

\maketitle

\section{Introduction}
An exciting facet of research in modern condensed matter physics deals with the manifestation of topology in the electronic states. Topological insulators (TIs) are quantum states of matter, having a gapped bulk like a normal insulator, and hosting protected states at the insulator-vacuum boundary \cite{hasan_colloquium_2010,shankar_topological_2018, qi_topological_2011}. These states are theorized to be robust against disorder as a result of the accompanying symmetry protection \cite{kane_quantum_2005}. There has been significant interest in employing materials manifesting topological phases for applications in electronics featuring these dissipation-less edge states \cite{gilbert_topological_2021, xu_manipulating_2017, nadeem_overcoming_2021,ezawa_quantized_2013}. Particularly, 2D-Xene monolayers and monolayer transition metal dichalcogenides (TMDs) are prominent candidates for 2D TIs \cite{ezawa_monolayer_2015,jana_robust_2022,banerjee_robust_2022}, with applications in topological electronics \cite{gilbert_topological_2021}. \\\indent 
Although theorized to be robust against backscattering, there is a need for detailed investigations of the effects of disorder in these topological phases if they are to be used for any topological electronics applications. Previous works have studied such effects in various regimes. Gap-opening mechanisms facilitated via inclusion of electron-electron interaction in a many-body picture has been reviewed in 2D TIs \cite{balram_current-induced_2019, xu_stability_2006, chou_gapless_2018, wu_helical_2006} and 3D TIs \cite{baum_magnetic_2012, black-schaffer_spontaneous_2014}. Interestingly, the combination of electron-electron interactions and momentum-dependent spin polarization leads to a mean-field hamiltonian similar to a Zeeman coupling \cite{balram_current-induced_2019}, dubbed a `local magnetization' \cite{black-schaffer_spontaneous_2014}. The number of Kramer's pairs operational in the helical state also impacts the stability, as outlined in \cite{xu_stability_2006, wu_helical_2006}. Effects of magnetic impurities have been studied in \cite{chen_massive_2010, zhu_electrically_2011}. A general analysis of the effects on incoherent electromagnetic noise on the edge states is presented in \cite{vayrynen_noise-induced_2018}. \\\indent 
In this work, we employ the non-equilibrium Green’s function (NEGF) formalism \cite{datta_electronic_1995, camsari_non-equilibrium_2023} to study the effects of disorder and dephasing in 2D Xene-devices that are capable of hosting quantum spin Hall (QSH) and spin quantum anomalous Hall (SQAH) phases. The NEGF formalism provides a phenomenological computational framework for introducing different kinds of dephasing within tight-binding models \cite{golizadeh-mojarad_nonequilibrium_2007,Aritra,Praveen,Duse_2021}. Particularly, NEGF provides the possibility of introducing momentum dephasing, which relaxes momentum as well as phase, and spin dephasing, only relaxes spin, independently of each other \cite{camsari_non-equilibrium_2023}. We study the effects of momentum and spin relaxation and analyze in depth how the topological phase responds to dephasing via looking at the conductance quantization in the presence of disorder. We also compare the methods of NEGF enabled momentum dephasing and explicit addition of random potentials followed by sample averaging. The effect of Rashba spin-mixing on the robustness is studied, and a comparison of the robustness of the QSH and the SQAH phases is provided. We facilitate the analysis with plots of the bandstructures and wavefunction densities.
\\\indent 
The rest of the paper is organized as follows: In Sec. \ref{formalism}, we elaborate on the structure of the channel, the topological phases and the device under study, and outline the NEGF method for conducting the transport calculations. Next, we discuss in detail the effects of momentum and spin dephasing on the QSH and SQAH phases under various regimes in Sec. \ref{results}. We discuss the implications of the results and outline future work in Sec. \ref{conclusion}.
\begin{figure*}[t]
    \centering 
\includegraphics[width = \textwidth, height = 9.5cm]{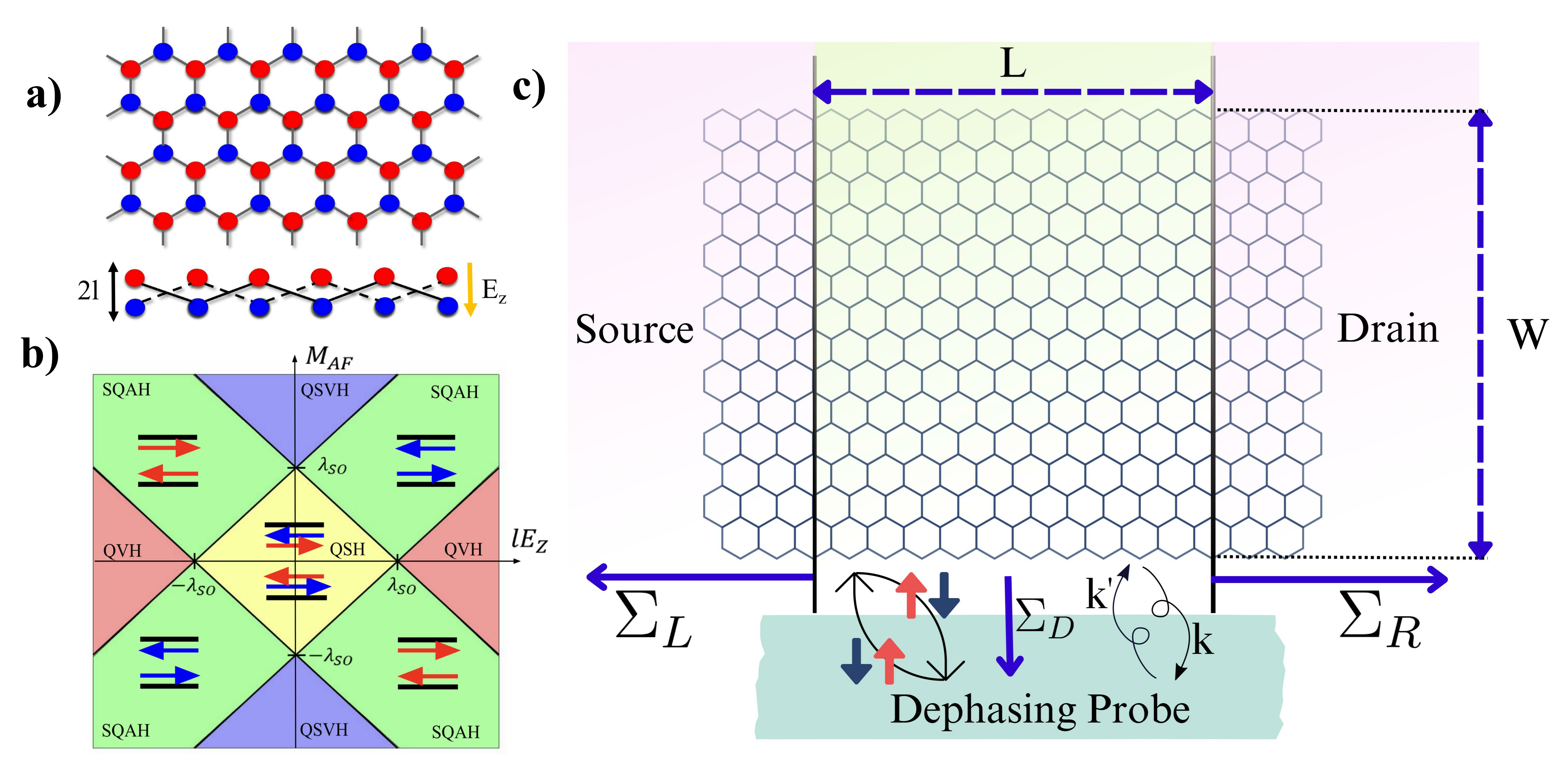}
    \caption{a) Top and side views of the Xene nanoribbon: Top view shows the honeycomb lattice structure, atoms from both the sublattices are depicted in red and blue colors. Side view of shows the buckled nature of the honeycomb lattice. The buckling height is given by $2l$. An out-of-plane electric field $E_z$ applied perpendicular to the nanoribbon induces a potential difference between the two sublattices, given as $V(A) = \Delta_z$ and $V(B)= -\Delta_z$, where $\Delta_z = lE_z$ (b) Phase diagram showing the possible topological phases of the device under the considered hamiltonian as a function of the spin-orbit coupling strength $\lambda_{SO}$, sublattice potential $\Delta_z = lE_z$, and the antiferromagnetic coupling strength $M_{AF}$ \cite{banerjee_robust_2022}
    (c) Device structure under consideration: The channel, hosting the topological phases, is connected to the source (left, $L$) and drain (right, $R$) contacts. Under the NEGF model, the contacts are modelled with self energies $\Sigma_{L/R}$, and dephasing is modelled through the phenomenological NEGF dephasing probem, modelled as $\Sigma_D$, which includes (i) spin dephasing, and (ii) momentum dephasing, shown as arrows. }
    \label{fig:fig1}
\end{figure*}
\section{Formalism and Setup}\label{formalism} 
\subsection{Device Structure and Hamiltonian}
We investigate the QSH and SQAH phases in the 2D buckled hexagonal lattice structure as shown in Fig. \ref{fig:fig1}(a). The top view of the hexagonal lattice shows the $A$ and $B$ sublattice points in red and blue respectively. The side view shows the buckled nature of the lattice with buckling height $2l$. The electric field $E_z$ applied perpendicular to the lattice creates a staggered sublattice potential $\Delta_z = lE_z$, with the net potential difference between the sublattice points being $2\Delta_z$. The low energy Bloch hamiltonian, is given as, \begin{multline}
    \mathcal{H}(\mathbf{k}) = \underbrace{\hbar v_f (k_x\tau_z\sigma_x + k_y\sigma_y)}_{\text{Nearest neighbour hopping}} + \underbrace{\lambda_{SO}\sigma_z\tau_z s_z}_{\text{Spin-orbit coupling}} + \\ 
    \underbrace{\lambda_R(\sigma_x s_y \tau_z  -  \sigma_y s_z)}_{\text{Rashba spin-mixing}} + \underbrace{\Delta_z \sigma_z}_{\text{Staggered sublattice potential}} + \\ 
    \underbrace{M_{AF}\sigma_z s_z}_{\text{Antiferromagnetic interaction}}
\end{multline}
where $\sigma_i$, $\tau_i$, and $s_i$ denote the pauli matrices in the space of the sublattice points $A$-$B$, the valley index $K$-$K'$ and the spin $\uparrow$ - $\downarrow$ for $i \in \{x,y,z\}$. The quantity $\hbar$ is the reduced Planck's constant and $v_f$ is the Fermi velocity given as $v_f = 3ta_0/2$, with $t$ being the nearest neighbour hopping and $a_0$ the side length of the hexagonal lattice. Moreover, $\lambda_{SO}$ is the strenght of the spin-orbit coupling, $\lambda_R$ denotes the Rashba spin mixing interaction, and $\Delta_z$ encapsulates the perpendicular electric field as discussed before. Lastly, the antiferromagnetic interaction, which can be realized via proximity coupling is quantified using the $M_{AF}$ term. The tight-binding hamiltonian used for numerical calculations can be found in \cite[Appendix~A]{jana_robust_2022}.\\\indent  
The interplay of the parameters $(\lambda_{SO},\Delta_z,M_{AF})$ results in the realization of various topological phases in the channel, which are adiabatically disconnected from each other. Figure \ref{fig:fig1}(b) elucidates these phases diagramatically. Notably, in the absence of antiferromagnetic interaction ($M_{AF} = 0$): for $\Delta_z < \lambda_{SO}$ the channel is in the QSH state with zero total Chern number and a non-zero spin Chern number, indicating helical edge states. With antiferromagnetic interaction ($M_{AF} \neq 0$): for $\Delta_z > M_{AF} - \lambda_{SO}$ the channel is in the SQAH phase with a non-zero total Chern number, possessing spin-polarized chiral edge states. 

\subsection{Transport Calculations}
We consider the following device structure as outlined in Fig. \ref{fig:fig1}(c): The channel is $L$ atoms long and $W$ atoms wide, and is sandwiched between the source and the drain terminals. All the proximity effects and electric fields are present in the channel region only. That is, the electric field, causing the sublattice potential $\Delta_z$ and the Rashba spin-mixing $\lambda_R$, along with the antiferromagnetic interaction $M_{AF}$ is considered zero outside the $L \times W$ channel region for the calculations below. \\\indent The NEGF method, is a means of attaching contacts to the Schr\"{o}dinger equation for performing quantum transport calculations \cite{datta_electronic_1995, camsari_non-equilibrium_2023}. Moreover, the NEGF method also enables the modelling of phase-breaking non-coherent processes withing the self-consistent Born approximation. We use the NEGF method along with the Landauer-Buttiker formalism for our numerical simulations. In the regime of coherent transport, the Retarded Green's function $G_R$ for the channel is defined as, 
\begin{equation}
    G^R(E) = ((E+\iota\eta)I - H - \Sigma_L - \Sigma_R)^{-1}
\end{equation}
where, $\eta$ is a small damping parameter, $H$ is the channel (tight-binding) hamiltonian, $I$ is the identity matrix, $\Sigma_{L(R)}$ are the self energies describing the left $(L)$ and the right $(R)$ contacts. The advanced Green's Function is defined as $G^A = (G^R)^{\dagger}$. Using the contact self-energies, the broadening matrices are defined as
\begin{equation}
    \Gamma_{L/R} = \iota(\Sigma_{L/R} - \Sigma^{\dagger}_{L/R})
\end{equation}
We get the spectral function, $A$ as 
\begin{equation}
    A(E) = \iota[G^R(E) - G^A(E)]
\end{equation}
The diagonal elements of the spectral function represent the local density of states. The electron correlation function matrix is given as
\begin{equation}
    G^n(E) = G^R(E)[\Sigma^{in}_L + \Sigma^{in}_R]G^A(E)
\end{equation} 
where, the in-scattering matrices are given as $\Sigma^{in}_{L/R}(E) = f_{L/R}(E)\Gamma_{L/R}(E)$ using the fermi functions $f_{L/R}$ of the contacts. The transmission from the left to the right contact at the energy $E$ is computed as, 
\begin{equation}
    T(E) = Tr(\Gamma_L G^R\Gamma_R G^n)
\end{equation}
The conductance is given as $G(E) = e^2/h \times T(E)$. \\\indent
\begin{figure}[t]
    \centering
    \includegraphics[width = 8cm, height = 12cm]{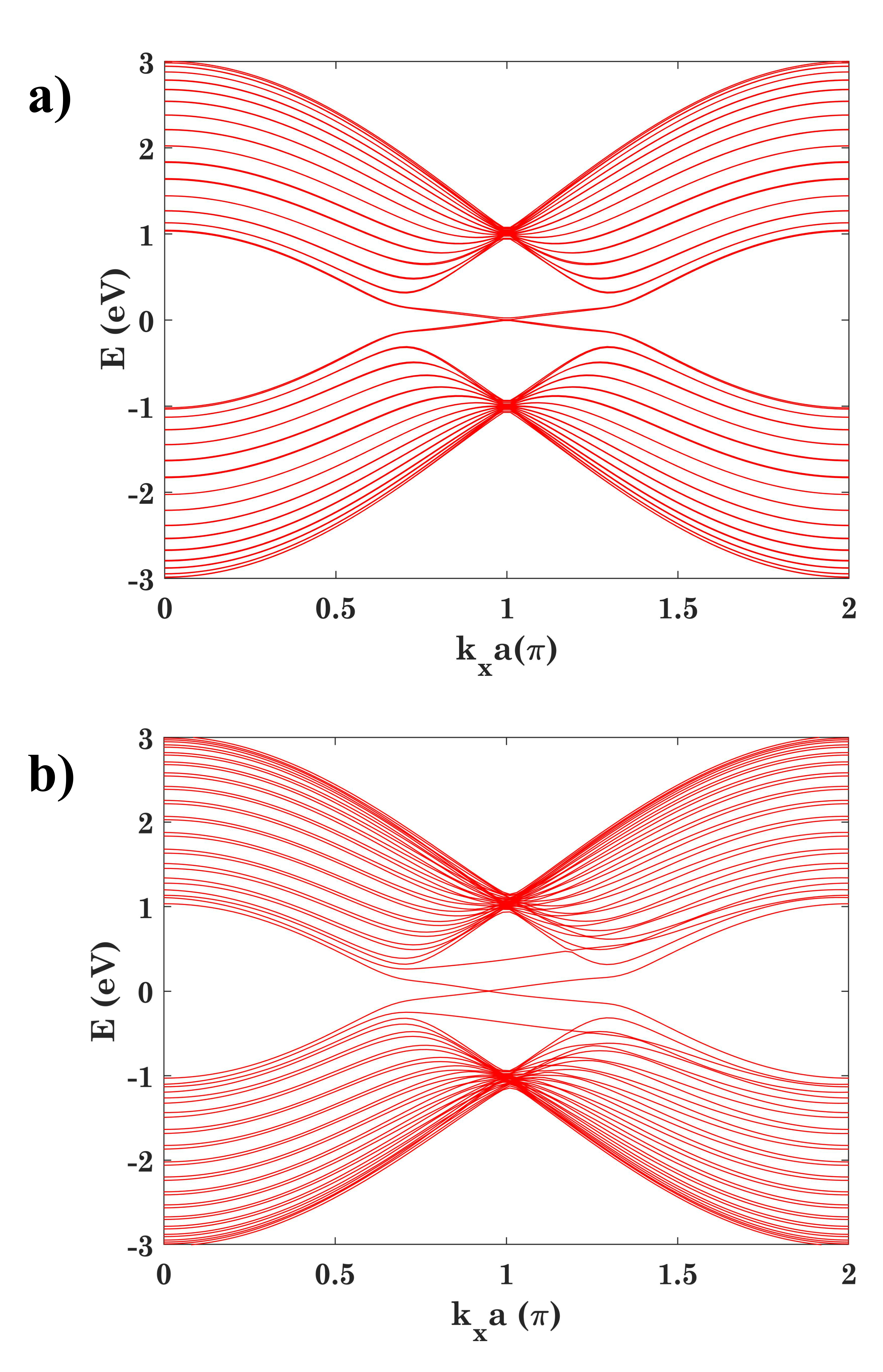}
\caption{Band structure diagrams for the (a) QSH phase with $\lambda_{SO} = 0.15eV$, $\Delta_z = 1meV$ and $\lambda_R = 0$; the (b) SQAH phase with $\lambda_{SO} = 0.15eV$, $M_{AF} = 0.201eV$ $\Delta_z = 0.2eV$, and $\lambda_R = 0$. The bandgap for both the phases is $0.298 eV$.}    \label{fig:my_label}
\end{figure}
\begin{figure*}[t]
    \centering
    \includegraphics[width = 18cm, height = 12cm]{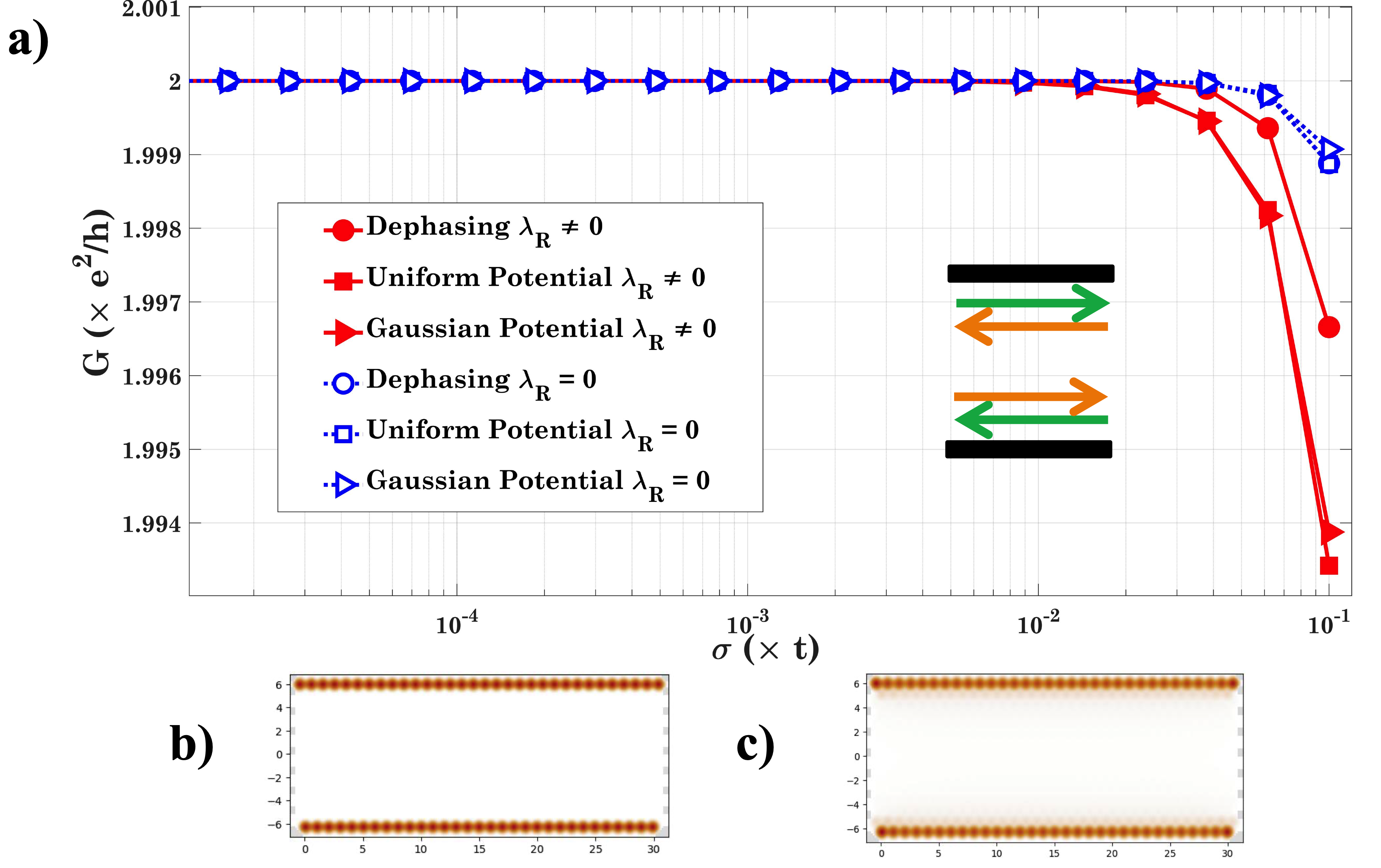}
\caption{Effect of momentum dephasing on the QSH phase: We employ the NEGF dephasing model with verying dephasing strengths $D$  add IID disorder potentials of variance $\sigma^2$ in the channel, with the relation $D_0 = \sigma^2$ (a) Conductance for the cases $\lambda_R = 0$ (blue, dotted) and $\lambda_R = 15meV$ (red, filled) for various cases of disorder and dephasing; (b) Wavefunction density over the channel for a pristine QSH phase (c) Wavefunction density over the channel for momentum dephasing with $D_0 = 0.1t^2$}
    \label{fig:fig3}
\end{figure*}

\begin{figure*}[t]
    \centering
    \includegraphics[width = 18cm, height = 12cm]{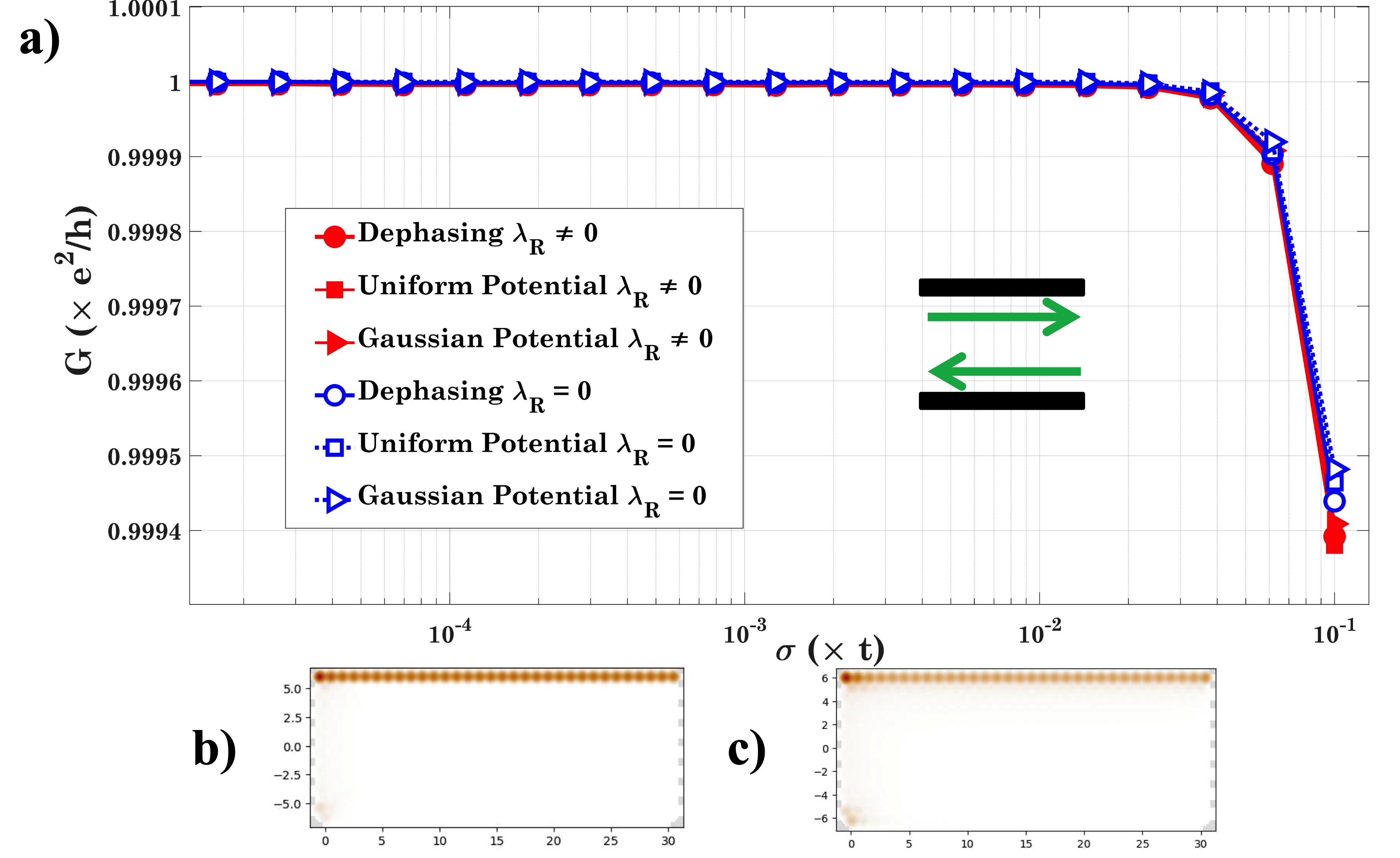}
\caption{Effect of momentum dephasing on the SQAH phase: We employ the NEGF dephasing model with verying dephasing strengths $D_0$ and add IID disorder potentials of variance $\sigma^2$ in the channel, with the relation $D_0 = \sigma^2$ (a) Conductance for the cases $\lambda_R = 0$ (blue, dotted) and $\lambda_R = 15meV$ (red, filled) for various cases of disorder and dephasing; (b) Wavefunction density over the channel for a pristine SQAH phase (c) Wavefunction density over the channel for momentum dephasing with $D_0 = 0.1t^2$}
    \label{fig:fig4}
\end{figure*}

Using the NEGF method, one can simply add non-coherent phase-breaking processes using a ficticious self energy $\Sigma_D$ and in-scattering matrix $\Sigma^{in}_D$ \cite{golizadeh-mojarad_nonequilibrium_2007}. 
In this work, we concern ourselves with two models: 
\begin{enumerate}
    \item Momentum relaxation: We have, 
    \begin{align}
    \Sigma_D(i,j) &= \bar{D}(i,j)G^R(i,j) \equiv g_1(G^R) \\ 
    \Sigma^{in}_D(i,j) &= \bar{D}(i,j)G^n(i,j)  \equiv g_1(G^n)
\end{align}
    with $\bar{D}(i,j) = D_0\times \delta_{i,j}$. This choice of $(\Sigma_D,\Sigma^{in}_D)$ relaxes both phase and momentum. Moreover, this can be related to random impurity potentials through the equation 
    \begin{equation}
        \bar{D}(i,j) \sim \langle U_D(i) U^*_D(j)\rangle
    \end{equation}
    where $U_D$ denotes the impurity potential. In the simulations, we use the NEGF dephasing model as well as explicitly add random potentials.
    \item Spin relaxation: We have, 
    \begin{align}
        [\Sigma_D](i,j) &= D_0\sum_{k \in \{x,y,z\}}s_k[G^R](i,j) s_k \equiv g_2(G^R)\\ 
        [\Sigma^{in}_D](i,j) &= D_0\sum_{k \in \{x,y,z\}}s_k[G^n](i,j) s_k \equiv g_2(G^n)
    \end{align}
    where $[G^R](i,j)$ and $[G^n](i,j)$ denote $2 \times 2$ spin sub-blocks. This only relaxes the spin and conserves the momentum.
\end{enumerate}
Including both coherent and phase breaking processes, we summarise the calculations with the following four equations,
\begin{align}
    G^R &= ((E+\iota\eta)I - H - \Sigma_L - \Sigma_R - \Sigma_D)^{-1} \\ 
    G^n &= G^R(E)[\Sigma^{in}_L + \Sigma^{in}_R + \Sigma^{in}_D]G^A(E) \\ 
    \Sigma_D &= g(G^R) \\ 
    \Sigma^{in}_D &= g(G^n)
\end{align}
where $g \in \{g_1,g_2\}$ depends on the dephasing model under study. Numerically, these equations are solved self consistently under a stochastic approximation scheme.
\\\indent
We use the \texttt{KWANT} \cite{Groth_2014} package in python as the playground for all the numerical simulations. The lattice structure is defined using \texttt{KWANT}, which enables access to the self-energies of the contacts. Thereafter, the subsequent analysis is done using the NEGF method outlined above.
\section{Results and Discussion}\label{results}
In this section, we report the results of our numerical simulations. A Xene nanoribbon of length $L = 30$ atoms and width $W = 13$ atoms is used for the analysis, unless otherwise stated. Moreover, the parameter values used for the QSH and SQAH phases are summarized in Table \ref{tab:tab1}. Unless explicitly stated, the parameters in this table are used for the calculations. We first analyze the effects of momentum dephasing and random potential disorder, and then move on to spin dephasing.


\subsection{Momentum Dephasing \& Disorder}
\begin{table}[t]
    \centering
    \hspace{.1cm}%
    \begin{tabular}{c|r|r}
        \hline
        Parameter       & Value       & Units \\
        \hline
        \hline
        $\lambda_{SO}$       & $0.15$       & $eV$ \\
        \hline
        $\Delta_z$ & $1$ & $meV$ \\\hline 
        $\lambda_R$ & $\{0,15\}$ & $meV$ \\\hline  
        $M_{AF}$ & $0$ & $meV$  
    \end{tabular}%
    \hspace{.1cm}%
\quad
\begin{tabular}{c|r|r}
        \hline
        Parameter       & Value       & Units \\
        \hline
        \hline
        $\lambda_{SO}$       & $0.15$       & $eV$ \\
        \hline
        $\Delta_z$ & $0.2$ & $eV$ \\\hline 
        $\lambda_R$ & $\{0,15\}$ & $meV$ \\\hline  
        $M_{AF}$ & $0.201$ & $eV$ 
    \end{tabular}%
    \hspace{.1cm}%
    \caption{Parameter values for the numerical calculations performed for the QSH case (left table; with $\Delta_z < \lambda_{SO}$) and the SQAH case (right table; with $\Delta_z > M_{AF} - \lambda_{SO}$). These parameters are used unless otherwise stated. Further, a hopping value of $t = 1eV$ is used.}
    \label{tab:tab1}
\end{table}
We study the effects of disorder-induced momentum dephasing on the QSH and the SQAH phases. We add momentum depahasing using the NEGF model discussed in Sec. \ref{formalism} as well as random impurity potentials to the channel and average over hundreds of configurations. This acheives the purpose of studying the effect of dephasing on the topological phases, as well as stressing that the NEGF dephasing model is a viable alternative to potential averaging for studying disorder in tight-binding systems, where the latter can be computationally expensive. We add potential disorder under the independent and identically distributed (IID) regime, viz., the potential at each site is independent. Thus, we have $\langle U_D(i) U^*_D(j)\rangle = \langle U_D^2\rangle \delta_{i,j}$. Two models for the distribution are used: (i) Gaussian, $U \sim \mathcal{N}(0,\sigma^2)$ and (ii) Uniform $U \sim \text{unif}[-W,W]$; where, $\mathcal{N}(\mu, \sigma^2)$ denotes a Gaussian distribution with  mean $\mu$ and variance $\sigma^2$, $\text{unif}[-W,W]$ denotes the uniform distribution with support on $[-W,W]$, and $X \sim p(\cdot)$ denotes a random variable with $X$ with the distribution $p(\cdot)$. Since the magnitude of $D_0$ in the dephasing model depends on the variance of these distributions, we maintain, 
\begin{equation}\label{eq:dephrel}
    \sigma^2 = \frac{W^2}{12} \text{  and } ~~~ D_0 = \sigma^2,
\end{equation}
for consistency of the strength of disorder across the two potential-averaging and the dephasing method. Note that the variance of a $\text{unif}[-W,W]$ random variable is $W^2/12$.
\begin{figure*}[t]
    \centering
\includegraphics[width = 18cm, height = 8cm]{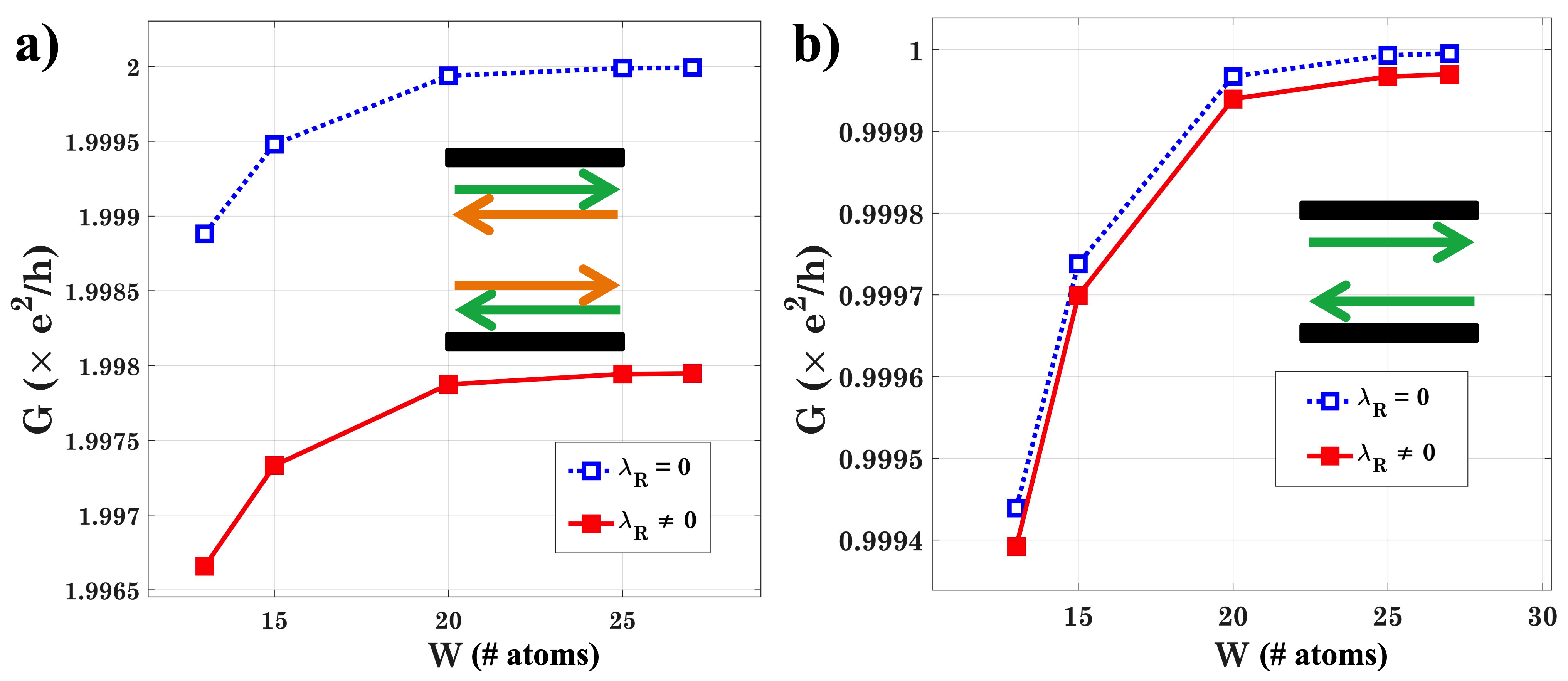}
\caption{Conductance for the cases $\lambda_R = 0$ (blue, dotted) and $\lambda_R = 15meV$ (red, filled) under momentum dephasing with $D_0 = 0.01t^2$ as a function of the width $W$, measured as the number of atoms, for the (a) QSH case and the (b) SQAH case}    
\label{fig:fig5}
\end{figure*}
\\\indent The effect of momentum dephasing and disorder potentials on the conductance of the QSH phase is outlined in Fig. \ref{fig:fig3}(a). We plot the conductance as a function of increasing disorder, quantified by the $\sigma$. Moreover, for the dephasing curves, the relation given in \eqref{eq:dephrel} is followed. We also show the wavefunction plots of the clean QSH phase as well as QSH under dephasing with $D_0 = 0.1t^2$.  We note that, for $\lambda_R = 0$, the QSH phase is starkly immune to disorder and dephasing with a worst-case decrease in conductance of $\sim 0.05\%$. The three models agree for QSH with $\lambda_R  = 0$. If we turn on the Rashba interaction, $\lambda_R \neq 0$, we note a worst case conductance drop of $0.3\%$. While still a miniscule decrease, it is to be noted that it is nearly a order higher than the previously noted decrease in conductance without Rashba mixing. Moreover, we note that for high disorder, the impurity potentials show worse conductance than the dephasing model. Further, the wavefunction density plots for the clean QSH channel and QSH under dephasing of strength $D_0 =0.1t^2$, both with $\lambda_R = 0$ are shown in Fig. \ref{fig:fig3}(b) and (c) respectively. In these, one observes a uniform discolouration under disorder, and the decay of the edge states into the bulk is visible. The plot for the dephased case with $\lambda_R \neq 0$ is similar, and not shown. 
\\\indent
The conductance response for the SQAH case is shown in Fig. \ref{fig:fig4}(a). We observe that for both the $\lambda_R$ cases the three models agree reasonably. The fact that conductance for the $\lambda_R \neq 0$ case is lower than that for $\lambda_R = 0$ at high disorder is much less pronounced than for the conductance response of QSH. The wavefunction plots are qualitatively the same as QSH, shown in Fig. \ref{fig:fig4}(b) and (c), and show discolouration under disorder, but show a lesser decay into the bulk. The worst case conductance drop is $0.06\%$, which is similar to the QSH case without Rashba mixing. 
\\\indent  
It is to be noted that the wavefunction plots only show the density for currents flowing from the left contact to the right contact, and hence, only one edge is illuminated for SQAH and both for QSH.
\\\indent \vspace{-3mm}
\begin{figure}[H]
    \centering
    \includegraphics[width = 7cm, height = 7cm]{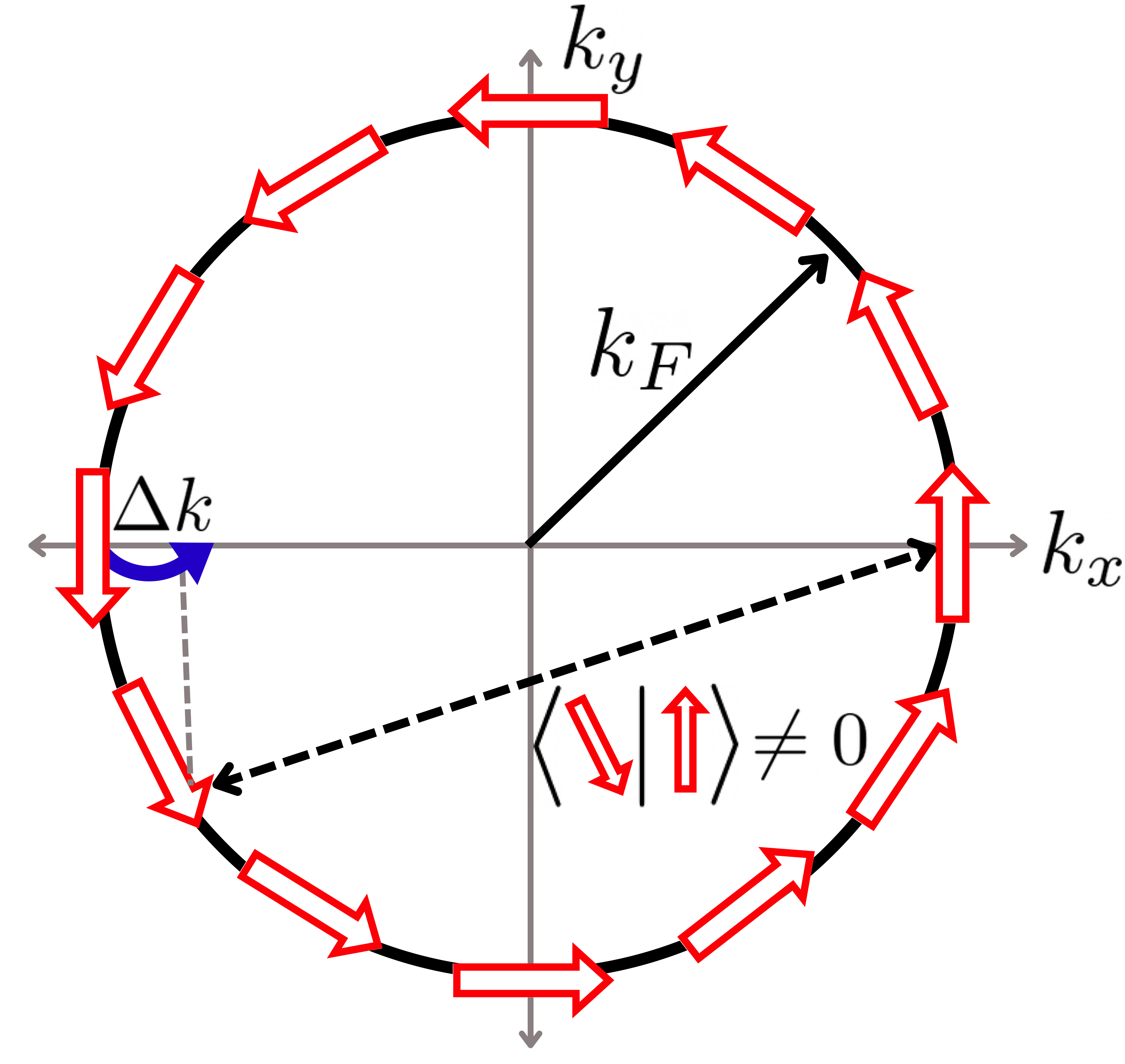}
\caption{Depiction of degradation caused by momentum relaxation in the presence of Rashba spin mixing interaction in the QSH phase: The Fermi surface of radius $k_F$ of the 2D sample along with the spin-texture caused by the Rashba interaction is shown. A momentum relaxation in a spin-down-left-mover, of order $\Delta k$, results in non-zero overlap with the spin-up-right-mover, thereby facilitating spin-flipping.}    \label{fig:rashbafig}
\end{figure}
\begin{figure*}[t]
    \centering
    \includegraphics[width = 18cm, height = 10cm]{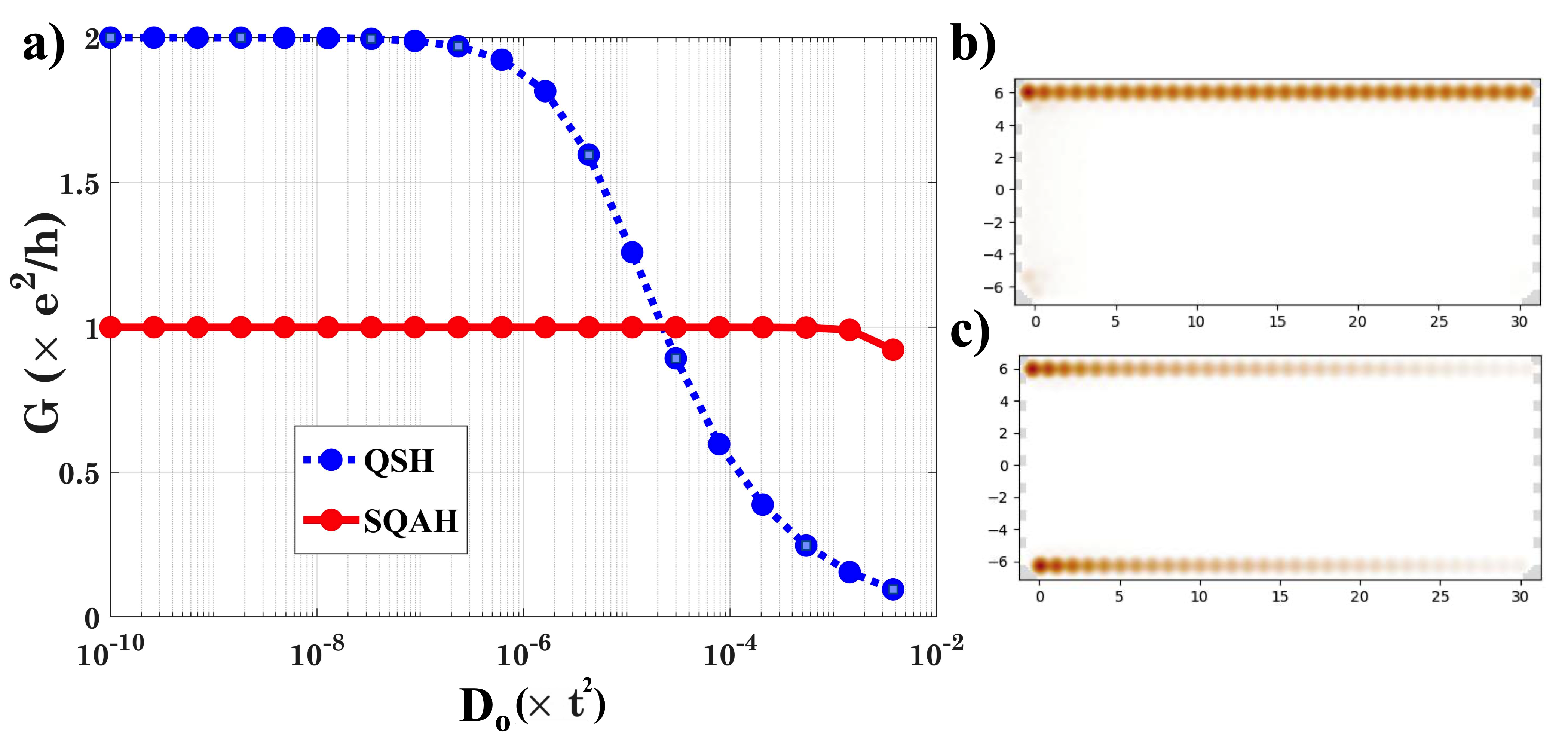}
    \caption{Effect of spin dephasing on topological phases: (a) Conductance as a function of the spin dephasing strength $D_0$ for the QSH (blue, dotted) and the SQAH (red, filled) phases (b-c) Wavefunction density across the channel for (b) the SQAH phase and (c) the QSH phase under spin dephasing of value $D_0 = 0.0014t^2$}
    \label{fig:fig6}
\end{figure*}
\begin{figure*}[t]
    \centering
    \includegraphics[width = 18cm, height = 12cm]{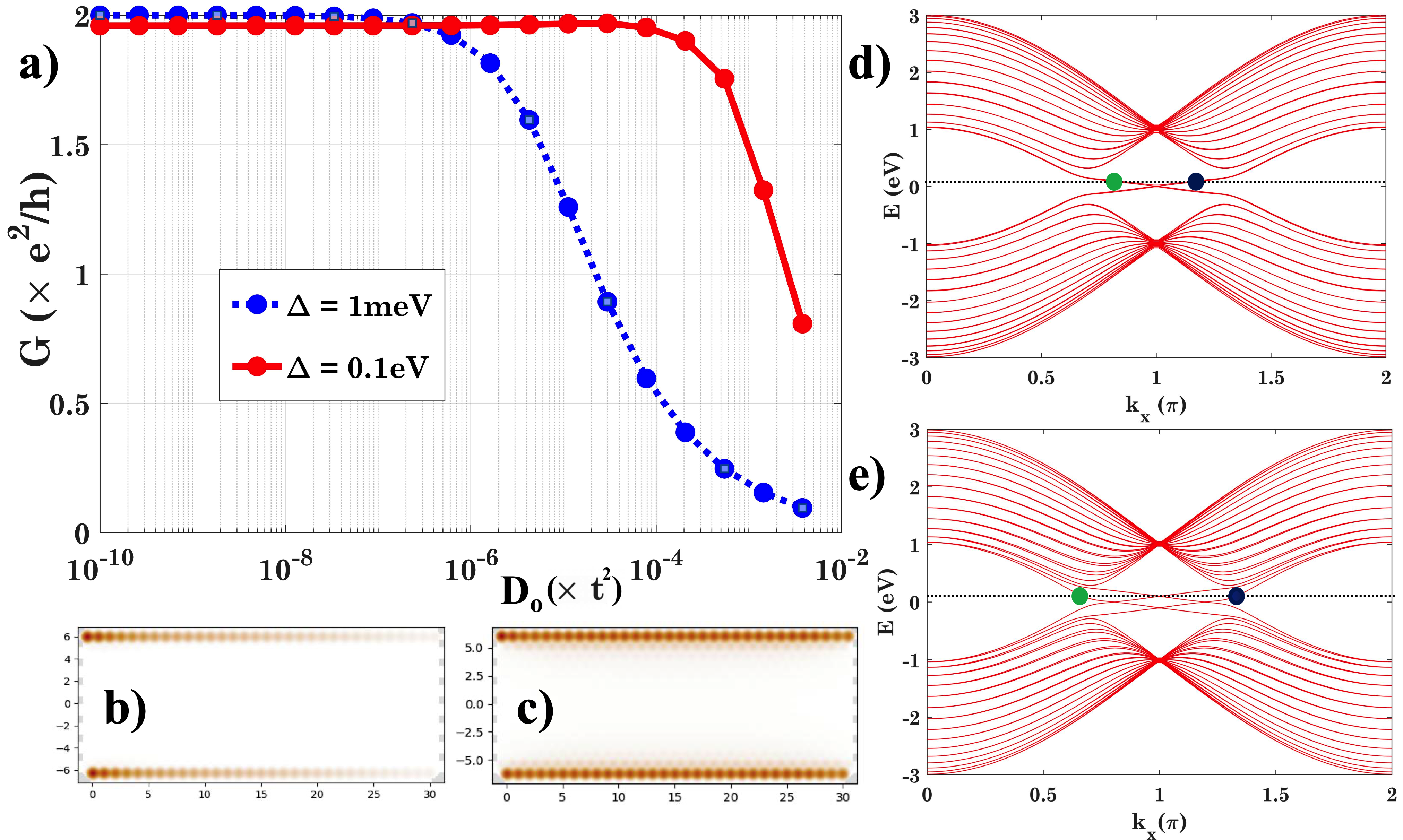}
    \caption{Interplay of the staggered potential $\Delta_z$ and spin dephasing in the QSH phase: (a) Conductance as a function of the spin dephasing strength $D_0$ for $\Delta_z = 1meV$ (blue, dotted) and the $\Delta_z = 0.1eV$ (red, filled) cases (b-c) Wavefunction density across the channel for (b) $\Delta_z = 1meV$ and (c) $\Delta_z = 0.1eV$ (d-e) Bandstructure diagram of QSH with (d) $\Delta_z = 1meV$ and (e) $\Delta_z = 0.1eV$. The dots on bandstructure diagram, for the same energy level (horizontal dotted line) illustrate momentum transfer required for spin-flipping.}
    \label{fig:fig7}
\end{figure*}
The different behaviours of Rashba spin-mixing in the two topological phases can be understood with the fact that the QSH phase is a \textit{helical liquid} with counter-propagating spin-locked edge modes on both sides of the sample, and the SQAH phase has \textit{chiral edge modes}, with only one spin direction forming the edge current. 
Introduction of the Rashba spin-mixing in an electronic systems causes a spin-texture to be formed at the Fermi surface \cite{Manchon_2015}, as shown in Fig. \ref{fig:rashbafig}. Now, consider the effect of momentum dephasing coupled with this spin-texture, as depicted in Fig. \ref{fig:rashbafig}. A stochastic change in momentum $\Delta k$ of a left-mover with spin down, can cause it to move to a different momentum state which is locked with another spin state. The latter now has the possibility of having a non-zero overlap with the right-mover up-spin state, hence facilitating backscattering. On the other hand, there is no such possibility for backscattering in the SQAH phase, as the mode is chiral and not helical. We elaborate more on this issue in the next subsection. \\\indent
Now, we study the effect of conductance under worst-case disorder of magnitude $D_0 = 0.01t^2$, which corresponds to $\sigma = 0.1t$, with the width of the channel $W$. The results are reported in Fig. \ref{fig:fig5}. For both the QSH and SQAH cases, one observes plateaus for $W > 25$. The pronounced drop in the QSH case owing to Rashba mixing can be seen clearly.
\subsection{Spin Dephasing}
We observed overall robustness of the topological phases under non-magnetic impurities captured by momentum dephasing and disorder potentials. Now, we will outline the effects of spin disorder on these phases using the NEGF spin dephasing model ($g = g_2$) as discussed in Sec. \ref{formalism}. \\\indent 
We show the effects of spin dephasing on the conductance of the QSH and SQAH phases in Fig. \ref{fig:fig6}(a). These are in stark contrast to the momentum dephasing results. We observe that the QSH phase is not robust to spin dephasing, and the QSH conductance drops to near zero for large dephasing values. Whereas, the SQAH phase remains robust and shows minimal drop in conductance even for large dephasing values. The wavefunction plots, shown in Fig. \ref{fig:fig6}(b-c) for SQAH and QSH respectively, support these findings, and show a decay of the QSH wavefunction across the length of the channel. The SQAH wavefunction appears to be faded, but is not decaying with the length. Note that spread of the wavefunction into the bulk is absent here, as compared to momentum dephasing. 
\\\indent 
Another interesting point is the role of $\Delta_z$ in the robustness of QSH to spin dephasing. For any value of $\Delta_z < \lambda_{SO}$ one can theoretically predict a QSH phase. But, is there any merit to choosing a particular value? We observe that, for higher $\Delta_z = 0.1 eV$ (less than $\lambda_{SO}$ to ensure the topological phase) the decay is much lesser than for a small $\Delta_z = 1meV$. The conductance response in this case is shown in Fig. \ref{fig:fig7}(a). The wavefunction density plots, shown in Fig. \ref{fig:fig7} (b) and (c) for $\Delta_z = 1meV$ and $\Delta_z = 0.1 eV$ respectively, illustrate this effect. We explain this through a careful look at the bandstructures for the two cases. These are plotted in Fig. \ref{fig:fig7}(d-e). Consider elastic scattering in spin from the green dot state to the blue dot state, at the same energy for the two cases of $\Delta_z$. For the $\Delta_z = 1meV$, the spin bands are overlapping and hence a small momentum transfer is required for the spin-flipping. For the $\Delta_z = 0.1 eV$ case, the spin bands move horizontally apart, and hence the spin flip requires a larger momentum transfer. Thus, for the same amount of disorder the latter case is much more robust to spin-flipping. \\\indent
We end this section with a general note about the spin dephasing model, and the results obtained.
With a magnetic impurity in the QSH case, time-reversal symmetry is broken and backscattering is possible between the counter-propagating modes at each edge. While in the SQAH case the edge modes are chiral -- there is no direct possibility of backscattering even in the presence of magnetic impurities. The behaviour of the edge modes thus leads to a clear contrast in the response of these phases under magnetic disorder, as was seen in Fig. \ref{fig:fig6}. This is qualitatively similar to how the spin-texture at the Fermi surface caused by Rashba mixing degrades the QSH case but not the SQAH case. It has been shown that spin flipping scattering can be caused by non-magnetic impurities \cite{balram_current-induced_2019} and particular forms of random Rashba spin-mixing interactions \cite{Zhang_2012}. Our spin dephasing results phenomenologically model a culmination of such effects, which would concern the robustness of such disorders in real devices.
\section{Conclusions}\label{conclusion}
This work was devoted to a computational exploration of topological robustness against various forms of dephasing. For this, we employed phenomenological dephasing models using the Keldysh non-equilibrium Green's function technique on a model topological device setup on a 2D-Xene platform. Concurrently, we also explicitly added disorder via impurity potentials in the channel and averaging over hundreds of configurations. To describe the extent of robustness, we quantified the decay of the conductance quantum with increasing disorder under different conditions. 
Our analysis showed that these topological phases are robust to experimentally relevant regimes of momentum dephasing and random disorder potentials. Moreover, we have pointed out a mechanism which accounts for the relatively worse performance of the QSH phase under momentum dephasing in the presence of Rashba interaction. We observed that the QSH phase break downs in the presence of spin dephasing, but the SQAH remains robust. Although, we showed that the QSH phase, under spin dephasing, performs better when the magnitude of the perpendicular electric field is higher; a mechanism for the same in terms of the bandstructures was outlined. The SQAH phase showed stark robustness under all the dephasing regimes, and can be potentially employed in realistic device structures for topological electronics applications.

\section*{Acknowledgements}
The authors wish to acknowedge insightful discusions with Supriyo Datta. The author BM wishes to acknowledge the support by the Science and Engineering Research Board (SERB), Government of India, Grant No. CRG/2021/003102, and the Ministry of Human Resource Development (MHRD), Government of India, Grant No. STARS/APR2019/NS/226/FS under the STARS scheme. The authors SM and BM acknowledge support of the Dhananjay Joshi Foundation under an Endowment to IIT Bombay. 



\nocite{*}
\bibliography{Dephasing_paper_bib}

\end{document}